# Bulk nanocrystalline Al-Mg-Y alloys with amorphous grain boundary complexions display high strength and compressive plasticity

Tianjiao Lei [1,2,6], Esther Hessong [1,6], Brandon Fields [1], Raphael Pierre Thiraux [1], Daniel S. Gianola [3], Timothy J. Rupert [1,4,5,*]

[1] Department of Materials Science and Engineering, University of California, Irvine, CA 92697, USA

[2] Metallurgical and Materials Engineering, University of Alabama, Tuscaloosa, AL 35401, USA

[3] Materials Department, University of California, Santa Barbara, CA 93106, USA

[4] Department of Materials Science and Engineering, Johns Hopkins University, Baltimore, MD 21218, USA

[5] Hopkins Extreme Materials Institute, Johns Hopkins University, Baltimore, MD 21218, USA

[6] These authors contributed equally to this work.

* Corresponding Author Email: tim.rupert@jhu.edu

## Abstract

Although nanocrystalline alloys regularly exhibit high strengths, their use in structural applications often face challenges due to sample size limitations, unstable microstructures, and the limited ability to plastically deform. The incorporation of amorphous grain boundary complexions has been proposed to address these issues, by simultaneously stabilizing nanocrystalline grain structures for scale-up processing and improving alloy toughness. In the present study, the mechanical behavior of bulk nanocrystalline Al-Mg-Y is examined with macroscale compression testing, probing a length scale that is relevant to real-world structural applications. Bulk samples were fabricated via a simple powder metallurgy approach, with different pressing temperatures and times employed for consolidation in order to investigate microstructural and property evolution. All of the specimens contained primary face-centered cubic Al and secondary $Al_4C_3$ and $Al_3Y$ phases, with the $Al_3Y$ particles exhibiting two populations of small equiaxed and larger elongated particles. Appreciable plasticity was measured along with high ultimate stresses over 800 MPa due to the presence of amorphous grain boundary complexions. Microstructural characterization of fracture surfaces revealed that the area fraction of dimpled regions increased with longer hot-pressing time. Most importantly, the elongated $Al_3Y$ particles formed regular cellular patterns with increasing hot-pressing time, delaying shear localization and significantly enhancing plasticity. The hierarchy present in the microstructure of the Al-Mg-Y alloy, from amorphous grain boundary complexions to secondary phases, gives rise to excellent bulk mechanical properties, which are attractive for structural applications.

## 1. Introduction

Nanocrystalline materials exhibit superior properties compared to their coarse-grained counterparts with one of the best-known examples being exceptionally high strength. However, the majority of synthesis techniques for these materials, such as magnetron sputtering, high pressure torsion, and equal-channel angular pressing, only allow for limited sample size [1,2,3] because nonequilibrium approaches are usually needed to prevent grain coarsening. For example, Azabou et al. [4] used severe plastic deformation to produce nanocrystalline Al alloys by applying high pressure of up to 7 GPa on mechanically milled powders at room temperature. Although these authors obtained relative density as high as ~97-98% and an average grain size of 71 nm, the sample thickness was only 4 mm. To scale up sample size, one common approach is combining high temperature, high pressure, and/or pulsed direct current for consolidation in order to remove pores between particles. Park et al. [5] employed hot extrusion and hot isostatic pressing on cryo-milled powders to produce a bulk nanocrystalline Al-Mg alloy. While the as-milled powders possessed nanosized grains, the final bulk sample exhibited large micrometer-sized grains due to recrystallization and grain growth during consolidation. To avoid grain growth, consolidation within a shorter period and/or under lower temperature is needed. Spark plasma sintering is a technique where pulsed direct current and uniaxial pressure are applied simultaneously to the sample, to reduce the required processing time [6] and achieve high heating rates of up to 1000 °C/min [7]. Ye et al. [8] used this technique to fabricate a bulk 5083 Al alloy by heating powders to 350 °C at a rate of 1.8 °C/sec followed by a hold period at that temperature for only 120 seconds under a pressure of 80 MPa. Despite achieving a sample with full density, microstructural characterization revealed that abnormal grain growth occurred with a volume fraction of coarse grains of ~10%. These coarse grains were observed to form on the outer surface of the powders, due to electric discharge and high current density at these regions. Therefore, fabricating bulk samples with nanosized grains remains challenging.

Recently, a new concept of designing nanocrystalline alloys via engineering amorphous grain boundary complexions was proposed [9,10], which is able to achieve both small grain size and bulk sample size. Amorphous complexions are phase-like disordered grain boundaries which form below bulk eutectic temperatures and are in local equilibrium with adjacent grains [11]. These complexions are usually a few nanometers thick and can significantly enhance mass transport due to the high excess free volume of the disordered phase, leading to solid-state activated sintering [12]. For instance, Gupta et al. [13] compared the sintering behavior between pure W and W doped with 0.5-1.0 at.% Ni. All samples were sintered at 1400 $^{o}$C for 2 h, with the pure W showing only ~1% densification while the W-Ni alloy exhibited more than 20% densification to a relative density of ~89%. Subsequent transmission electron microscopy (TEM) and Auger electron spectroscopy revealed the existence of amorphous complexions well below the bulk eutectic temperature in the W-Ni specimen [13]. In addition to activated sintering, these features can enhance thermal stability because of their lower free energy at elevated temperatures, with the degree of stability depending on solute elements. Lei et al. [14] examined the effect of different solute elements on the retention of amorphous complexions and the stabilization of grain sizes by comparing three binary Al alloys (Al-Mg, Al-Ni, and Al-Y). Y was identified to be the most efficient stabilizer of amorphous complexions, as these features in Al-Y were retained even for very slow cooling conditions, giving rise to equiaxed nanosized grains even after a hot pressing temperature of ~92% of the melting temperature of pure Al. Since amorphous complexions begin to form at a homologous temperature of 0.6-0.85 [11] and are stable at high temperatures, nanocrystalline alloys can exhibit small grain size even after exposure to high temperatures due to the formation of these features. Schuler et al. [15] observed that the grain size of nanocrystalline Ni-W was ~90 nm after being annealed at 900 ºC yet was only ~55 nm after annealing at 1100 ºC (corresponding to a homologous temperature of ~0.8), proving that there is a unique high temperature stability window. Complexion engineering has also been used to create bulk nanocrystalline pieces with both full density and nanosized grains in Cu-rich and Al-rich alloys [16,17].

Nanocrystalline alloys also typically have limited plasticity. For example, Wang et al. [18] observed that the failure strain of a nanocrystalline Cu processed via surface mechanical attrition treatment was still only ~2% under quasi-static tensile testing. Possible reasons for the low ductility include the ease of crack nucleation and unstable crack growth, resulting from a large volume fraction of high-energy grain boundaries which facilitate intergranular fracture propagation [19]. Ma et al. [19] further proposed that the early failure may be preceded or promoted by extremely localized plastic deformation. The onset of the shear localization is related to an instability of strain hardening originated from a lack of dislocation storage within small grains. Because dislocation generation and movement are largely inhibited, nanosized grains exhibit grain boundary-mediated deformation mechanisms [20] rather than dislocation-mediated mechanisms which are common in their coarse grain counterpart [21]. Therefore, strategies for improving plasticity while retaining the high strength of nanocrystalline materials are needed, especially for bulk specimens.

The present study investigates the bulk compressive plasticity of an Al-rich alloy, Al-Mg-Y, which was successfully fabricated into centimeter-sized pellets via grain boundary complexion engineering, as in previous studies [17]. Bulk pellets were fabricated via hot pressing consolidation of mechanically alloyed powders under different temperatures and durations to examine (1) microstructural evolution under different processing conditions and (2) the correlation between microstructure and deformation behavior. Secondary phases formed, including $Al_4C_3$ and $Al_3Y$, with the $Al_3Y$ particles exhibiting two populations: small equiaxed and larger elongated particles. For the shortest hot-pressing time, samples failed in a brittle manner by shattering into small pieces. For longer hot-pressing time, samples began to show appreciable plasticity, along with exceptional strengths greater than any commercially-available alternative. Analysis of fracture surfaces revealed that the elongated $Al_3Y$ particles began to form a cellular network with increasing hot-pressing time, which effectively delayed shear localization and further enhanced plasticity. The higher hierarchy of the microstructure, comprised of via mixture of amorphous grain boundary complexions as well as intermetallics with different structures, shapes and sizes gives rise to increased plasticity without significant loss of strength, which provides insights to mitigate the strength-plasticity tradeoff of lightweight structural alloys.

## 2. Materials and Methods

To synthesize bulk nanocrystalline Al-Mg-Y samples, powders of elemental Al (Alfa Aesar, 99.97%, -100+325 mesh), Mg (Alfa Aesar, 99.8%, -325 mesh) and Y (Alfa Aesar, 99.6%, -40 mesh) were first mechanically alloyed for 10 h to obtain a nanocrystalline solid solution. The dopant concentration was 2 at.% for both Mg and Y in the alloy and milling was conducted in a SPEX SamplePrep 8000M high-energy ball mill. Hardened steel vial and milling balls were used, and the ball-to-powder weight ratio was 10:1. In order to prevent excessive cold welding, 3 wt.% stearic acid was added as a process control agent. The milling process was conducted in a glovebox filled with Ar gas at an $O_2$ level <0.05 ppm to avoid oxidation. After milling, the alloyed powders were transferred into a ~14 mm inner diameter graphite die set and then consolidated into cylindrical bulk pellets using an MTI Corporation OTF-1200X-VHP3 hot press consisting of a vertical tube furnace with a vacuum-sealed quartz tube and a hydraulic press. The powders were first cold pressed for 10 min under 100 MPa at room temperature to form a green body and then consolidated under a hot-pressing temperature ($T_{HP}$) of 585 °C and a pressure of 100 MPa for 1, 3 , 6 , or 10 h. The heating rate used to reach the target $T_{HP}$ was 10 °C/min and the pellets were naturally cooled down to room temperature after pressing, which typically took more than 4 h. The dimensions of all bulk samples were ~1 cm in height and 1.4 cm in diameter. Readers are referred to Ref. [17] for more details on the consolidation process.

For quasi-static bulk compression testing, cylinders with 3 mm diameter and 6 mm height were sectioned via electrical discharge machining from the consolidated pellets. Compression tests were conducted at an initial rate of 0.001 $s^{-1}$ on an Instron 5985 frame equipped with a 250kN load cell, with strain calculations obtained from sample displacement. At least four and up to six specimens were tested for each condition in order to check for consistency. For porosity investigation, both optical microscopy and computed tomography (CT) scans were employed. Optical micrographs were taken on the cross section of each sample, which was first ground with SiC grinding paper down to 1200 grit and then auto-polished with monocrystalline diamond suspension down to 0.25 μm. To examine whether large pores formed within materials, scans were performed using a Xaria 410 Versa CT scanner (Zeiss, USA) with a voltage of 80 kV and a power of 15 W achieving a resolution of 13.6 μm. All scans were post-processed using ImageJ and a bandpass filter was applied to highlight the porosity. No visible pores were observed from the CT measurements for any of the samples investigated here.

Microstructural characterization was carried out using a variety of techniques. First, X-ray diffraction (XRD) measurements were conducted to identify phases, calculate phase fraction, and measure grain size. XRD was performed using a Rigaku Ultima III X-ray diffractometer with a Cu Kα radiation source operated at 40 kV and 30 mA and a one-dimensional D/teX Ultra detector, with analysis conducted using an integrated powder X-ray analysis software package (Rigaku PDXL). Scanning electron microscopy (SEM) imaging for fracture surface examination and backscattered electron (BSE) imaging for intermetallic distribution investigation were performed in an FEI Quanta 3D FEG dual-beam SEM/Focused Ion Beam (FIB) microscope. Bright-field (BF) and high-angle annular dark field (HAADF) scanning transmission electron microscopy (STEM) were used to examine grain sizes and precipitates inside of a JEOL JEM-2800 S/TEM, which was operated at 200 kV and equipped with a Gatan OneView IS camera.

## 3. Results

### 3.1. Porosity and Microstructure of Bulk Nanocrystalline Al-Mg-Y Alloys

Figure 1(a) shows a representative bulk pellet ~1 cm in both diameter and height corresponding to a 1 h hot-pressing ($t_{HP}$ = 1 h) consolidation, where no visible pores were observed on the surface. Figures 1(b) and (c) show cylindrical specimens that were extracted with electrical discharge machining for bulk compression testing. The dimensions of all cylinders (3 mm in diameter and 6 mm in height) were chosen so that bending and buckling can be avoided [22]. Porosity and microstructure were first examined, with Figures 1(d)-(g) presenting cross-sectional optical images for all hot-pressing times explored here. Full density (>99.8%) was confirmed for all conditions. Moreover, these images show different contrast, suggesting the formation of secondary phases. The concentrations of both solute elements in the present study are 2 at.% each, which is above the solubility of Mg in Al at room temperature but below the solubility of Mg in Al at temperatures higher than 160 ºC. In contrast, this composition is above the solubility limit of Y in Al at both room and higher temperatures [23,24]. Therefore, it is possible that intermetallic phases formed during the hot-pressing or cooling process of consolidation. Secondary phases may also form due to the stearic acid ($C_{18}H_{36}O_2$) added during ball milling, which adds impurities to the system.

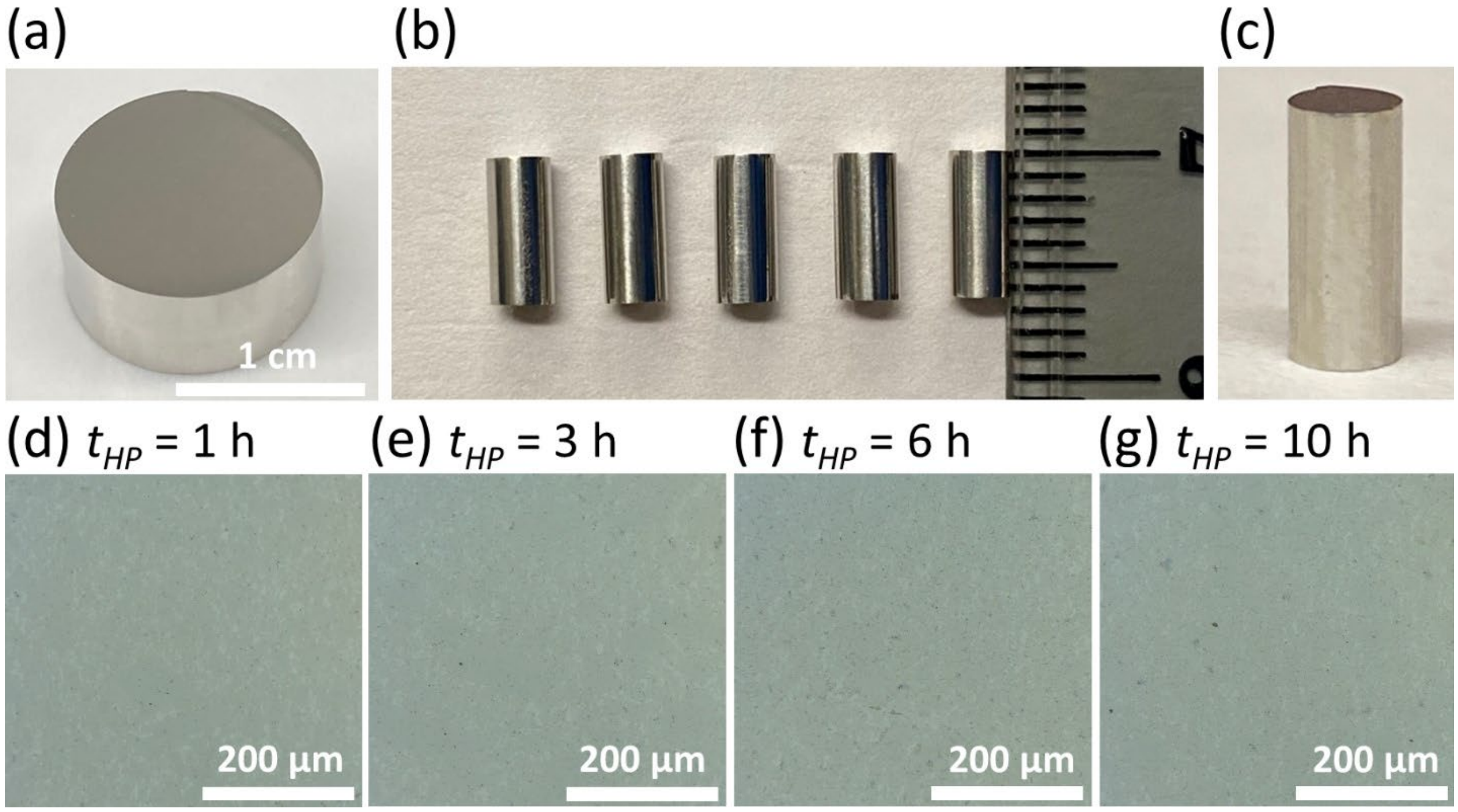

**Figure 1. (a) A pellet of Al-Mg-Y after hot pressing at 585 ºC for 1 h. (b),(c) Cylindrical specimens for compression testing. All samples had dimensions of 3 mm in diameter and 6 mm in height, with an aspect ratio of 2. (d)-(g) Cross-sectional optical micrographs for samples with hot-pressing time ($t_{HP}$) from 1 to 10 h, respectively. No obvious pores were observed for any of the consolidated samples.**

Figure 2(a) shows XRD scans for all hot-pressing conditions, where peak positions and intensities are similar. In addition to the face-centered cubic (FCC) Al-rich matrix phase, $Al_3Y$ and $Al_4C_3$ peaks emerged. Both secondary phases have a trigonal crystal structure with a space group of R-3m [25,26]. With increasing $t_{HP}$, the peak width of the Al matrix becomes narrower, suggesting an increasing grain size. Figure 2(b) shows the evolution of the average matrix grain size obtained from the XRD scans as a function of hot-pressing time. For $t_{HP}$ = 1 h, the matrix grains exhibited an average size of 44 ± 4 nm, while the value increased to 107 ± 29 nm as $t_{HP}$ reached 10 h. These grain sizes are consistent with TEM observations shown in Figures 2(d) and (h). Kinetics of normal grain growth follow [27,28,29]:

$$D - D_0 = Kt^n, \quad (1)$$

where $D$ and $D_0$ correspond to grain size for a heat treatment time of $t$ and 0 s, respectively, $K$ is a constant dependent on composition and temperature, and $n$ is the grain growth exponent. In the present study, $D_0$ is approximately equal to the grain size of as-milled powders, 30 nm, yielding a growth exponent of 0.52, close to the value in the classical work by Burke and Turnbull [30]. The average grain size for $t_{HP}$ = 10 h is 107 nm, close to the value for $t_{HP}$ = 6 h, 102 nm. The retention of these small grains after exposure to a homologous temperature of 0.92 points to excellent thermal stability, which is partly due to nanoscale precipitates located at grain boundaries, as indicated by yellow arrows in Figures 2(e) and (i). These nanoscale precipitates possess an elongated shape and therefore are termed as "nanorods". Moreover, the edge and the interior of the nanorods exhibit different contrast in these HAADF-STEM micrographs, suggesting variations in chemical composition at these two locations, as detailed below.

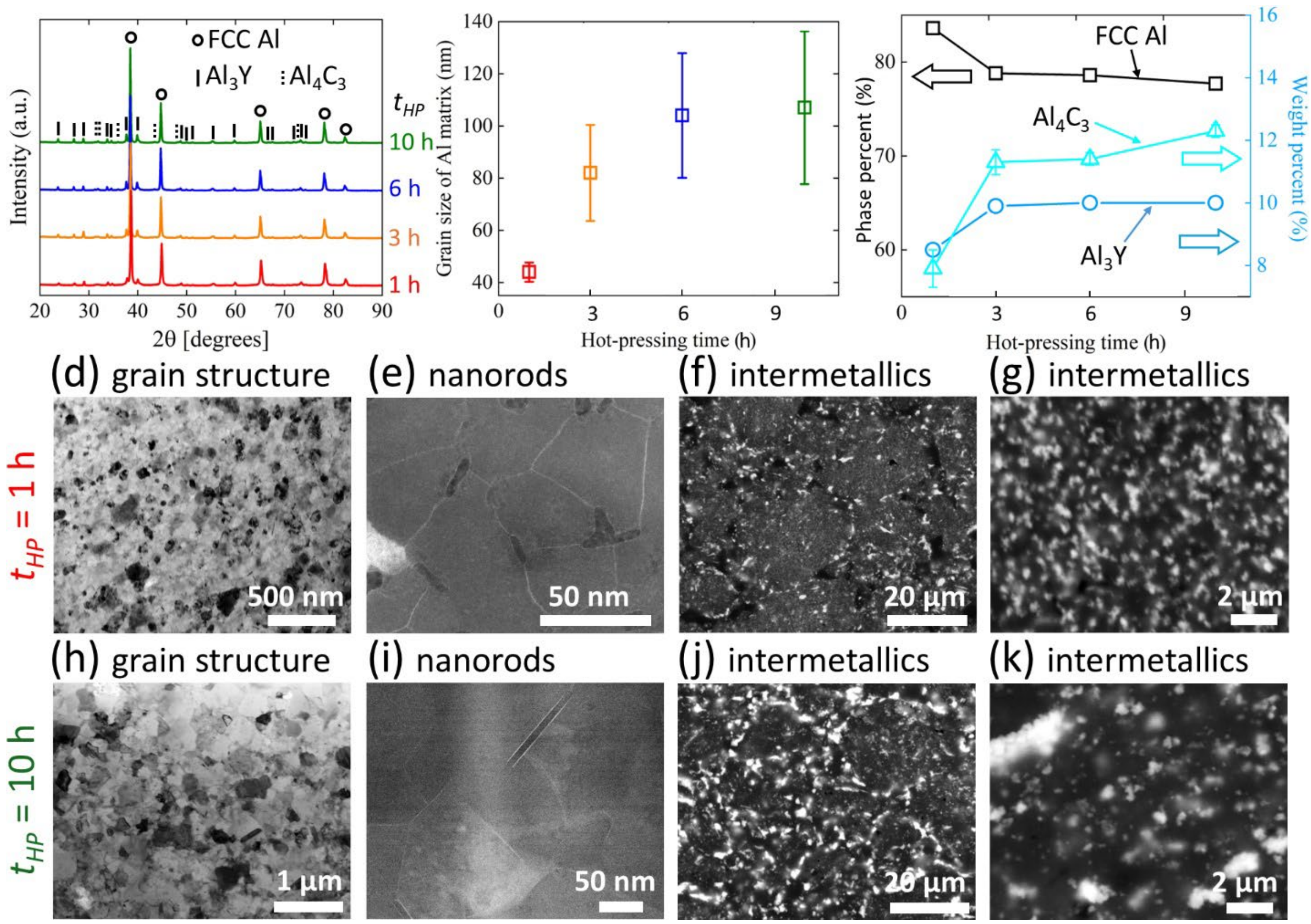

**Figure 2. Microstructural characterization of Al-Mg-Y. (a) XRD scans for all hot-pressing conditions show evidence of three phases: FCC Al, $Al_3Y$, and $Al_4C_3$. (b) Grain size of the FCC Al matrix calculated from XRD as a function of hot-pressing time. (c) Weight percent of the three phases obtained from XRD data as a function of hot-pressing time. (d) and (h) Low-magnification BF-STEM micrographs showing the overall grain structure for $t_{HP}$ = 1 h and 10 h, respectively. (e) and (i) HAADF-STEM micrographs showing nanorod precipitates that originate at grain boundaries for $t_{HP}$ = 1 h and 10 h, respectively. (f) and (j) Low-magnification BSE micrographs showing two populations of intermetallic phases for both conditions: large, elongated particles and small, equiaxed particles. (g) and (k) Zoomed views showing that the spatial distribution of equiaxed intermetallic phases for $t_{HP}$ = 1 h and 10 h, respectively.**

Figure 2(c) shows the weight fractions of the three phases as a function of hot-pressing times. For the shortest pressing time, the weight percent of $Al_3Y$ and $Al_4C_3$ are 8.5%, and 7.9%, respectively. As $t_{HP}$ increased to 3 h, the weight percent of the intermetallic phase increased to 9.9% and that of the carbide phase increased to 11.3%. With longer $t_{HP}$, the fractions did not vary significantly, reaching respectively 10% and 12.3% for $Al_3Y$ and $Al_4C_3$ when $t_{HP}$ = 10 h. To examine the size, morphology, and distribution of the secondary phases, both HAADF-STEM and BSE were employed. For the shortest hot-pressing time, nanorod precipitates that are a few nanometers wide and ~50 nm long was observed at almost every grain boundary, as mentioned before. The interior of these precipitates was identified to be $Al_4C_3$ in our previous study, while the brighter edges of these precipitates is due to segregation of Mg and Y to the matrix-precipitate interfaces [17]. As $t_{HP}$ increased to 10 h, the carbide precipitates coarsened to become ~10 nm wide and ~100 nm long (Figure 2(i)), while their location remains at the grain boundaries.

To examine the $Al_3Y$ phase, BSE imaging was used because this technique offers a larger survey area than TEM. Figures 2(f) and (j) present an overall view of the intermetallic phase for $t_{HP}$ = 1 h and 10 h, respectively, where two populations of particles emerged. The majority have an equiaxed shape as clearly shown in Figures 2(g) and (k), while large, elongated particles also formed. Similar equiaxed $Al_3Y$ particles were observed in Al-Ni-Y alloys fabricated by hot compacting and extruding gas-atomized amorphous powders by Vasiliev et al. [31]. These particles were characterized by a space group of R-3m and sizes of 50-300 nm in diameter. In another study of an Al-10 at.% Y system during solidification, the $Al_3Y$ phase was found to possess an elongated morphology with both widths and lengths on the order of tens of micrometers [32]. Since the equiaxed particles have smaller sizes than the elongated ones in both previous works and the present study, we hypothesize that the former may be still at the initial stage of particle growth. For $t_{HP}$ = 1 h (Figure 2(f)), the small equiaxed particles seem to aggregate in regions with sizes of ~20-30 µm, while the larger elongated particles sit at the perimeter of these regions. As $t_{HP}$ reached 10 h (Figure 2(j)), the size and volume fraction of both intermetallic populations increased significantly compared to $t_{HP}$ = 1 h. Most importantly, the elongated particles formed a regular cellular pattern on the order of ~10-20 µm, which enclosed the equiaxed particles with sizes of ~500 nm-1 µm (Figure 2(k)).

One important microstructural feature in the present Al-Mg-Y alloy is the existence of amorphous grain boundary complexions, with Figure 3 showing an example enclosed between yellow dashed lines and separating two crystalline grains (marked G1 and G2). The thickness of this complexion is ~2 nm, consistent with those observed in a nanocrystalline Al-Ni-Y alloy with similar processing conditions [14,17,33,]. The formation of amorphous complexions in the present Al-Mg-Y can be attributed to four key factors proposed in Ref. [34]: (1) positive segregation enthalpy, (2) limited solubility, (3) negative pair-wise mixing enthalpy, and (4) large atomic size mismatch. The first two factors promote segregation of solute elements to grain boundaries, while the latter two are essential to facilitate the formation of a disordered grain boundary structure. Amorphous complexions are stable at high temperatures (typically above 0.6-0.85$T_m$) and usually transition back to an ordered state at lower temperatures. Accordingly, previous studies on Cu-rich alloys required rapid quenching to retain amorphous complexions at room temperature, e.g., [16]. In contrast, in the present study, all Al-Mg-Y samples were naturally cooled down to room temperature with a cooling rate less than 1 ºC/s, yet amorphous complexions were still found throughout the microstructure. This indicates an exceptional kinetic stability of amorphous complexions in the present Al-Mg-Y system, which should be due to a combination of the large negative mixing enthalpy between Al and Y as well as the existence of multiple elemental species in the grain boundary, giving rise to a good glassing forming ability for the boundary region [34,35,36]. Large negative mixing enthalpy and multiple elements are critical rules for the formation of bulk metallic glasses, which exhibit amorphous nature [37]. The similar structural disorder in both bulk metallic glasses and amorphous grain boundary complexions gives rise to common criteria for their formation.

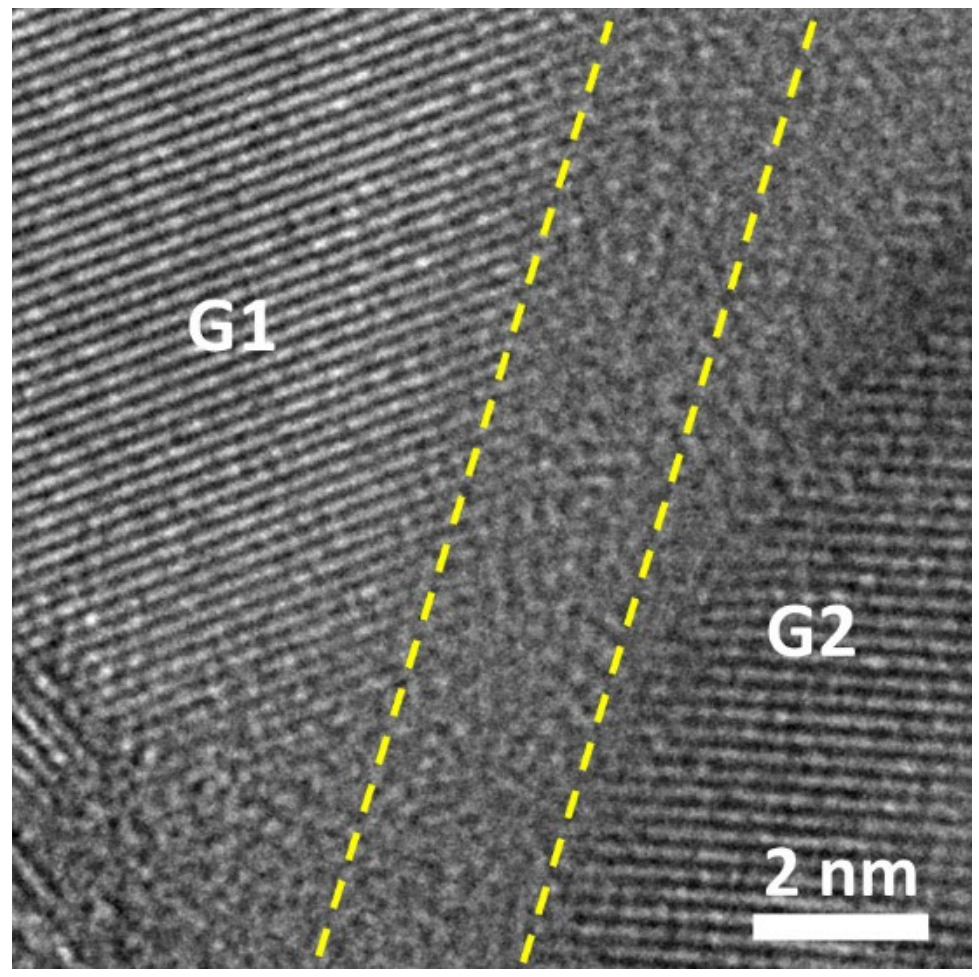

**Figure 3. High-resolution TEM micrograph of one representative amorphous grain boundary complexion in the present nanocrystalline Al-Mg-Y alloys.**

### 3.2. Mechanical Properties and Deformation Behavior

Figure 4 shows engineering stress-strain curves from bulk compression testing of the Al-Mg-Y samples. For the shortest hot-pressing time ($t_{HP}$ = 1 h), the specimens failed prior to obvious macroscopic yielding. Hence, all samples shattered into pieces before reaching a yield point and exhibited a fracture strength (~500 MPa) that is lower than the other specimens were measured. We note that this strength level is quite high for an Al alloy, yet the lack of plasticity diminishes the usefulness of any strength gains. The 1 h sample also had a noticeably lower Young's modulus, likely due to some limited internal porosity. Young's modulus can be related to porosity by the expression,

$$E = E_0 \times \exp(-\beta P) \tag{2}$$

where $E_0$ is the reference, fully dense value, $\beta$ is a constant ranging from ~3-4.5 and $P$ represents porosity [38]. While metallography and CT scanning did not identify any pores in the present work, it is possible that a limited numbers of scattered pores did exist and affected the mechanical behavior. As $t_{HP}$ increased to 3 h, Al-Mg-Y began to show measurable plasticity, with ultimate strengths higher than 800 MPa. With longer hot-pressing time, the plasticity increases further while the strength decreases. For $t_{HP}$ = 6 h, the average yield stress and strain-to-failure are 785 ± 6 MPa and 4.4 ± 0.4%, respectively. For $t_{HP}$ = 10 h, the average yield stress and strain-to-failure are 620 ± 5 MPa and 6.4 ± 1.5%, respectively. It is worth noting that the compressive yield strengths for both $t_{HP}$ = 6 h and 10 h are much higher than those of high-strength commercial Al alloys, e.g., Al7075. For example, El-Magd and Abouridouane [39] performed room-temperature compressive testing with a nominal strain rate of 0.001 $s^{-1}$ on an AA7075 alloy (5.9 wt% Zn, 2.4 wt% Mg, 1.5 wt% Cu) and reported a compressive yield strength of 420 MPa (labelled in Figure 4). The much lower yield strength in Ref. [39] is most likely due to larger matrix grain sizes, which are on the order of tens of micrometers in Ref. [39] versus ~80-100 nm in the present study. In addition to the excellent yield strength, the specific strength of the present Al-Mg-Y alloys is remarkable. Based on the alloy composition in Ref. [39], the estimated density of the AA7075 system is ~2.79 g/cm$^3$, while the density of the present Al-Mg-Y is ~2.74 g/cm$^3$. As a result, the-strength-to-weight ratio of Al-Mg-Y is 1.5-1.9 times higher than that of AA7075 alloys, which can potentially improve fuel efficiency for aerospace applications. Besides 7075 alloys, the present Al-Mg-Y alloys are even stronger than 7075 alloys reinforced with $Al_2O_3$ and $B_4C$ particles (with representative yield strengths being shown in Figure 4), where the ceramic reinforcements are expected to provide more strength [40,41]. This highlights the cooperative effect of all microstructural features, including nanosized grains, amorphous grain boundary complexions, nanorod precipitates, and various intermetallic particles, on strengthening the nanocrystalline Al-Mg-Y alloys, which can be easily processed using a simple and conventional powder metallurgy approach. Another advantage of the present Al-Mg-Y system over commercial Al7075 alloys and Al7074 reinforced with ceramics is their excellent stability, mainly due to the presence of amorphous complexions (Figure 3), allowing for a higher operation temperature regime. Besides excellent strengths, the microstructural features give rise to superior bulk plasticity in the present Al-Mg-Y system, as the plasticity for $t_{HP}$ = 6 h and 10 h can reach 5% and 8%, respectively. For nanocrystalline alloys, measurement of bulk plasticity is challenging, since samples with nanosized grains were rarely successfully fabricated into pieces that are large and dense enough for bulk compression testing. Khan et al. [42] performed a series of compaction and hot sintering procedures, including compaction under pressures >2 GPa and room temperature for 2 min, sintering under 900 MPa and 600-635 ºC for 1 h, sintering under 600 ºC for 11 h, and annealing under 600 ºC for 1 h, to fabricate bulk nanocrystalline Al and then performed quasi-static compression testing on cylindrical samples with diameters and lengths of ~1 cm. An average grain size of ~40 nm was achieved but the corresponding compressive plasticity was only ~2%, much lower than the plasticity of the Al-Mg-Y system studied here. For many reports on nanocrystalline Al alloys, plasticity is usually obtained from small-scale mechanical testing, such as nano- or micro-mechanical compressive testing [43,44,45], and therefore the reported values may not represent actual bulk plasticity.

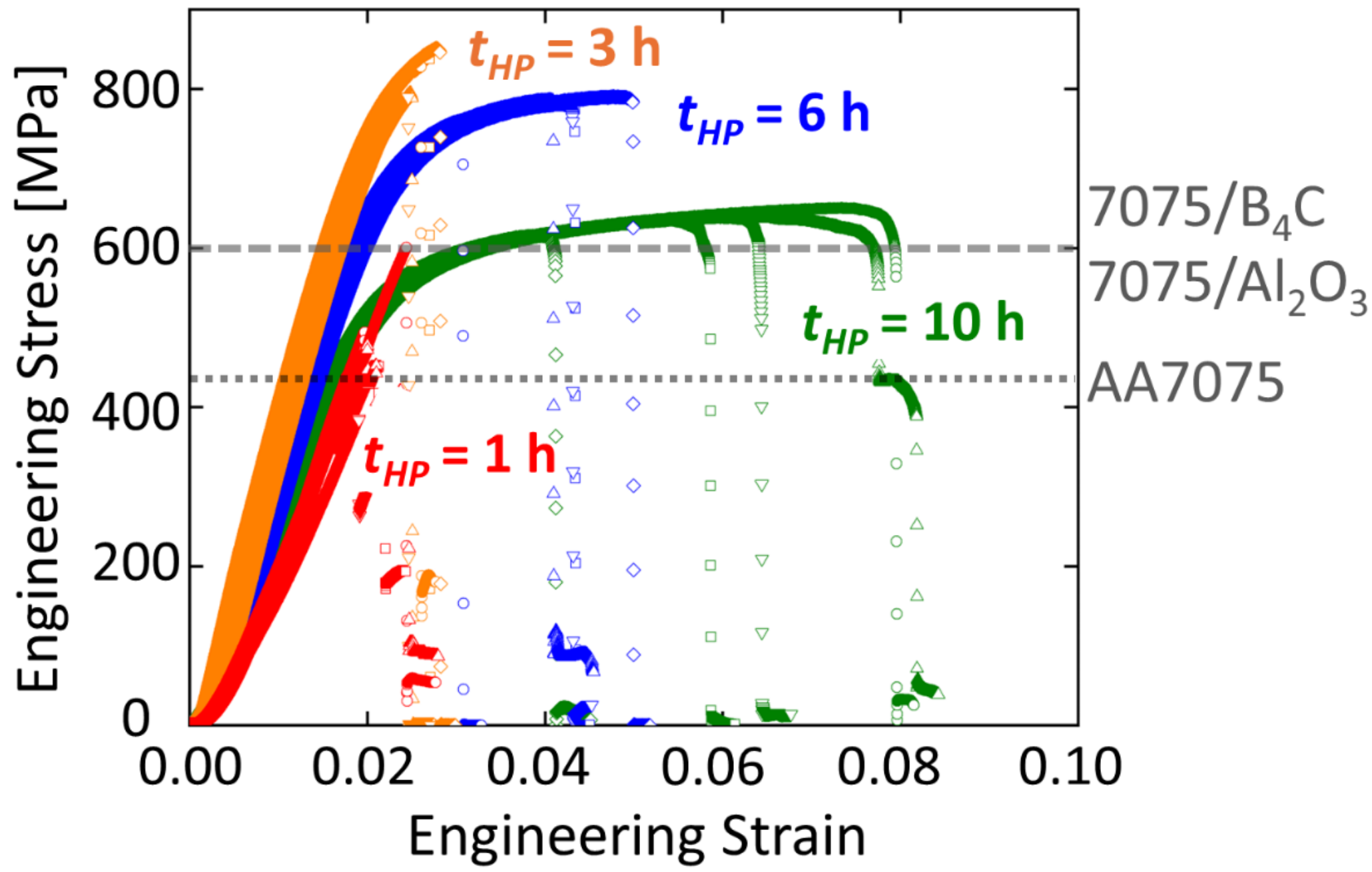

**Figure 4. Engineering stress-strain curves of bulk Al-Mg-Y samples after various hot-pressing times ($t_{HP}$), with dotted and dashed lines corresponding to yield strengths of commercial 7075 alloys and 7075 alloys reinforced with ceramics [39,40,41], respectively.**

After bulk compression testing, deformed samples were examined using SEM. Figure 5(a) presents a representative deformed sample for $t_{HP}$ = 1 h, with Figure 5(b) showing many pieces came off during testing. In addition, samples with the shortest hot-pressing period did not experience any stable flow and the vertical dashed lines in Figure 5(a) align very well with the sample side.

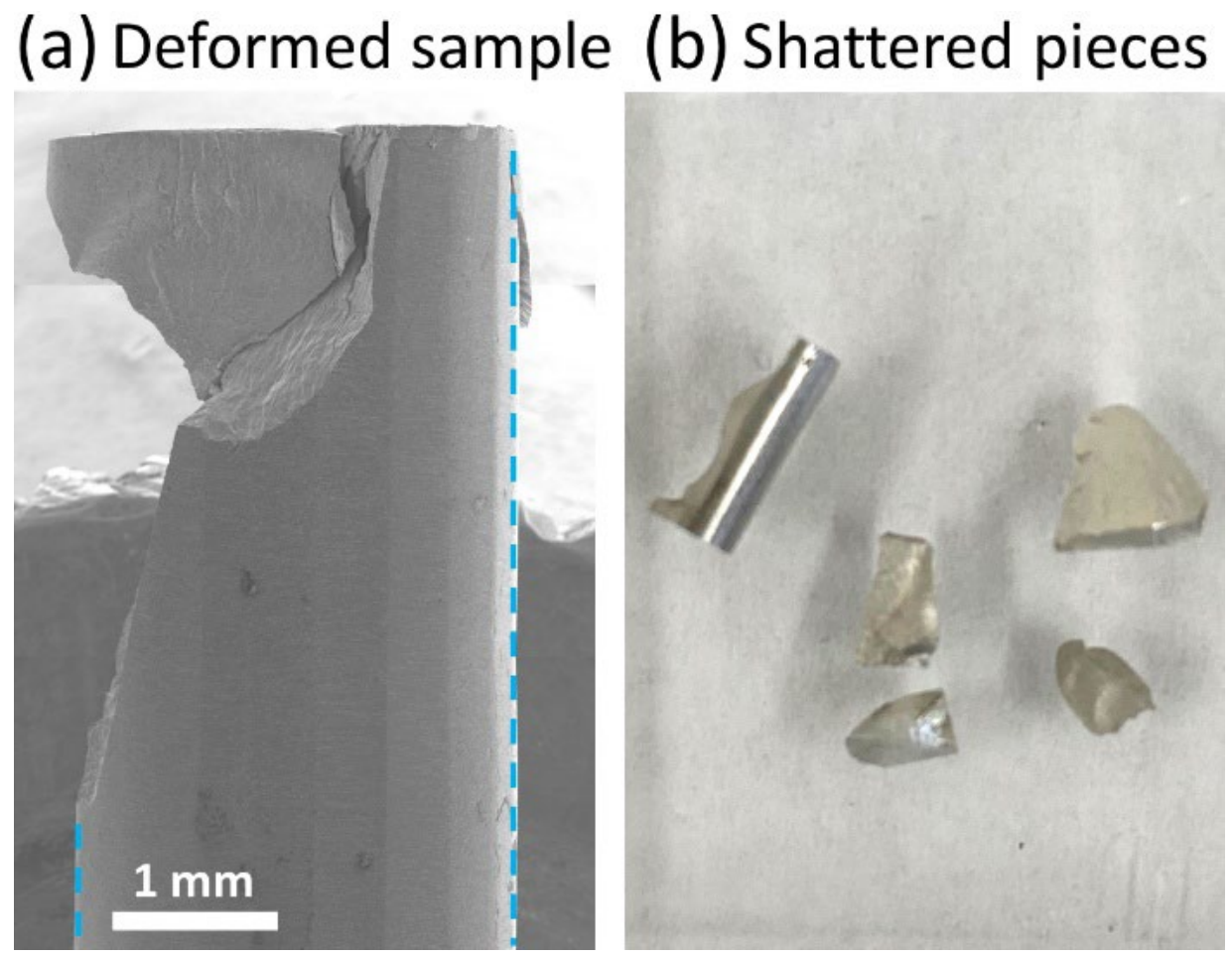

**Figure 5. (a) SEM micrograph of a representative deformed $t_{HP}$ = 1 h sample with brittle failure, and (b) photo of shattered small pieces after testing for the same condition.**

Figures 6(a) and (b) correspond to two samples after compression testing for $t_{HP}$ = 6 h, which clearly showed a different failure mode from $t_{HP}$ = 1 h. First, both samples with $t_{HP}$ = 6 h failed through a dominant shear band. Second, the samples with a longer consolidation time experienced slight barreling, evidenced by the difference between the straight dashed lines and sample surfaces in both Figures 6(a) and (b).

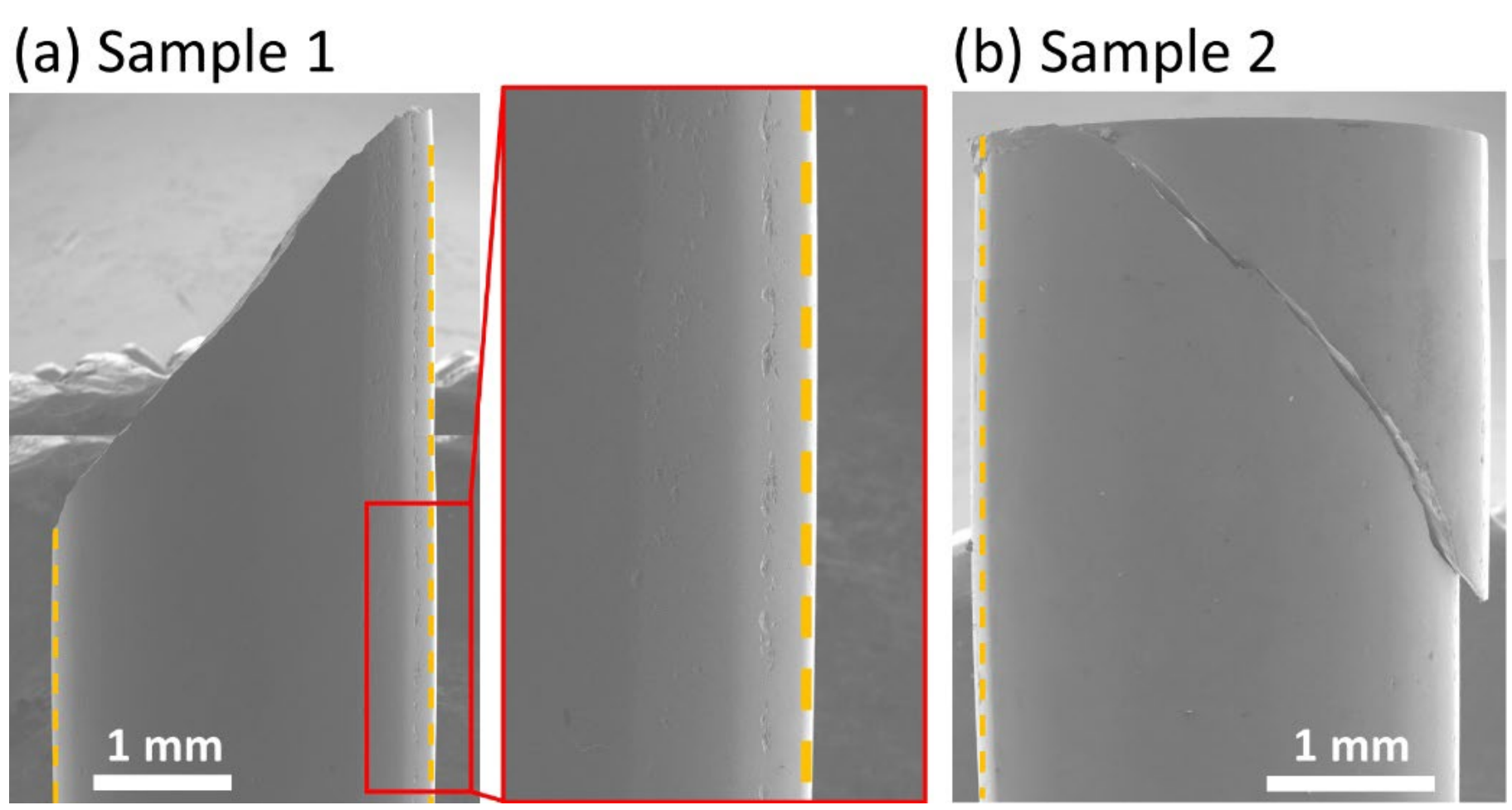

**Figure 6. (a)&(b) SEM micrographs of two samples corresponding to $t_{HP}$ = 6 h, which experienced stable plastic flow indicated by the slight barreling, which is shown as the difference between the straight dashed lines and the sample surface.**

### 3.3. Fracture Surface Morphology and Intermetallic Particle Distribution

To understand the different deformation behavior, fracture surfaces were examined and are presented in Figure 7. For $t_{HP}$ = 1 h, cleavage fracture occurred since the surface exhibited facets with different orientations (top panel of Figure 7(a)). River patterns with spacings of ~50 μm were also observed, further confirming the brittle fracture mode. This spacing is wider than the <10 μm value reported for a nanocrystalline tungsten fabricated via high-pressure torsion after dynamic compression testing [46]. Formation of river patterns is often associated with regions of weak atomic bonding [47]. In the present study, likely weak regions are interfaces between matrix and intermetallic phases, which could be nucleation sites for river pattern formation. Despite the macroscopically brittle behavior for $t_{HP}$ = 1 h, zoomed views showed a co-existence of both brittle and ductile fracture modes. For the brittle region, multiple shear bands were observed, while vein patterns formed in the ductile area (indicated by yellow arrows in the bottom panel of Figure 7(a)), suggesting local plastic deformation. Moreover, some vein patterns enclose sub-micrometer particles (with one example being marked by a red arrow), the size of which is similar to that of equiaxed $Al_3Y$ particles in Figure 2(g). Therefore, these inclusions are most likely intermetallic particles. Due to the limited area fraction of these plastic deformed regions, $t_{HP}$ = 1 h exhibited an overall brittle failure.

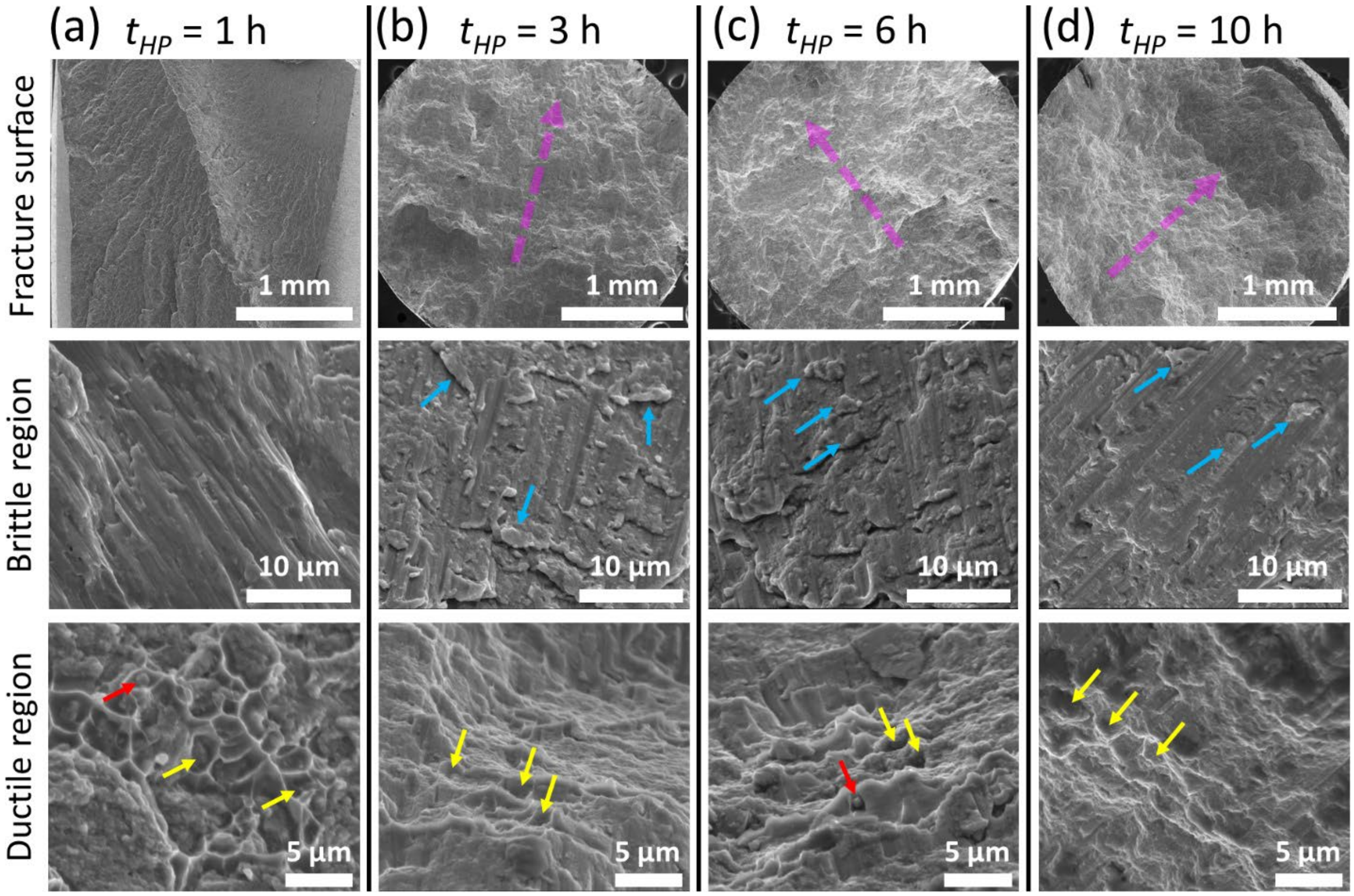

**Figure 7. The morphology of fracture surface for (a) $t_{HP}$ = 1 h, (b) $t_{HP}$ = 3 h, (c) $t_{HP}$ = 6 h, and (d) $t_{HP}$ = 10 h, where both brittle- (middle row) and ductile-type (bottom row) fracture were observed in local regions.**

Figures 7(b)-(d) show fracture surfaces corresponding to longer hot-pressing times (i.e., $t_{HP}$ = 3 h, 6 h, and 10 h) These samples which demonstrated macroscopic plasticity in the stress-strain curves all had similar fracture surfaces. All exhibit wavy patterns pointing in a certain direction (magenta dashed arrows), indicating that the fracture initiates at the tail of the arrow and then grows along the arrow direction. Zoomed views in the bottom panels reveal that these wavy patterns are dimples and examples are marked by yellow arrows on the ductile regions. For all conditions, the dimple sizes were found to be similar, 1-2 µm, or approximately 10 times the matrix grain size. This ratio is consistent with a previous study on tensile testing of electrodeposited nanocrystalline Ni alloys [48], where dimple sizes were about six to ten times of the average grain size. In Ref. [48], the dimples were observed to nucleate at grain boundaries and triple junctions,

suggesting transgranular slip and unaccommodated grain boundary sliding to be responsible for dimple formation. The present Al-Mg-Y alloy may have a similar dimple formation mechanism because of the nanocrystalline grain sizes (~30 nm in Ref. [48] versus 80-100 nm in the present study) and possibly the same dislocation-mediated plasticity mechanisms. For the ductile regions, inclusions with sizes of ~1 μm were occasionally observed within the dimples and one example is shown with a red arrow for $t_{HP}$ = 6 h. Based on the size and morphology, these inclusions are most likely to be equiaxed $Al_3Y$ particles. Inclusions have been shown to affect the dimple size and therefore mechanical behavior. For example, Schuh et al. [49] performed tensile tests on a CrMnFeCoNi alloy with different heat treatment and observed that the size of the dimples formed around brittle secondary phases was larger than that without inclusions. The larger dimples suggested that more plastic deformation occurred locally, giving rise to better macroscopic plasticity in the end. However, in the present study, dimples with and without inclusions did not possess obvious different sizes, possibly due to the small size of equiaxed intermetallic particles (Figure 2).

Regions with signs of locally brittle fracture for $t_{HP}$ = 3 h, 6 h, and 10 h are shown in the middle panels of Figures 7(b)-(d), respectively. In contrast to $t_{HP}$ = 1 h, the longer hot-pressing times exhibited small islands (marked by blue arrows) on the brittle fractured surface. For $t_{HP}$ = 3 h, a majority of the islands possessed an elongated shape with sizes of a few micrometers. Both characteristics are similar to that of large, elongated $Al_3Y$ particles. The same observations are also shown for $t_{HP}$ = 6 h and 10 h in Figures 7(c) and (d). Therefore, intermetallic particles play an important role in mechanical behavior of Al-Mg-Y, because they affect both the ductile deformation as inclusions within dimples and brittle deformation as small islands.

Figure 8 shows a comparison of the brittle fractured region and the intermetallic distribution for all hot-pressing conditions. The top, middle, and bottom rows present brittle fracture surfaces, distribution of intermetallic particles, and crack and shear directions, respectively. When $t_{HP}$ = 1 h, the fracture surface exhibited a river pattern with spacings ranging from ~20 μm to 50 μm, as outlined by yellow dashed curves in the bottom panel of Figure 8(a). This pattern is very similar to the network formed by the elongated $Al_3Y$ particles shown in the middle panel of Figure 8(a). A magnified view of the interior of the river pattern shows thin shear bands (top panel of Figure 8(b)), where BSE imaging revealed that only small equiaxed particles existed within this region (middle panel of Figure 8(b)). The formation mechanism of river patterns has been previously attributed to preferential crack pathways along weakly bonded interfaces [47]. Comparing equiaxed and elongated $Al_3Y$, the latter may have a weaker interface bonding with the matrix at the particle tip due to a higher stress concentration [50]. This is consistent with a previous study on a die-cast A357 Al alloy, where flake- and needle-shaped Al-Si-La particles gave rise to large stress concentrations that acted as nucleation sites for cracks and therefore led to premature failure [51].

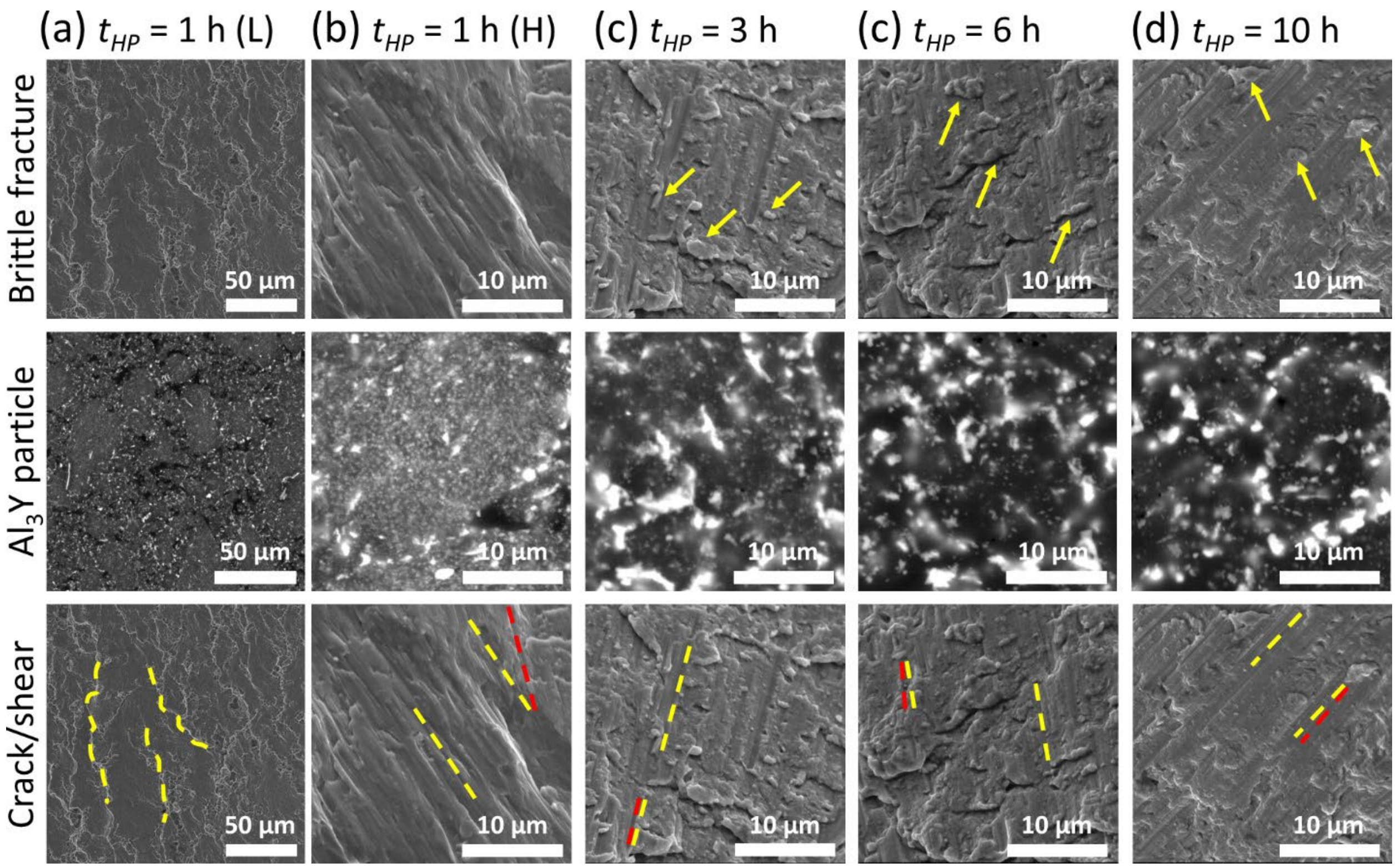

**Figure 8. Micrographs show the similarity between patterns on the brittle fracture surface and distribution of intermetallic particles for (a)&(b) 1-h, (c) 3-h, (d) 6-h, and (e) 10-h hot-pressing experiments. (a) and (b) show features at low (L)- and high (H)-magnification, respectively. The top and bottom rows present SEM images. Arrows in the top row marking small islands on the fracture surface, which began to form when $t_{HP}$ >= 3h, while dashed curves and lines in the bottom row outlining river patterns and shear directions, respectively. The yellow and red dashed lines in (b)-(d) represent different shear directions in two regions separated by intermetallic particles.**

For $t_{HP} \geq 3$ h, small islands were observed on the fracture surface, marked by yellow arrows in the top panels of Figures 8(c)-(e), and these islands shared a similar distribution with the large elongated $Al_3Y$ particles, which appear as brighter regions of a few micrometers in size in the BSE images, as shown in the middle panel of Figures 8(c)-(e). Moreover, as $t_{HP} \geq 3$ h, the elongated $Al_3Y$ particles began to form a cellular network. We hypothesize that this network provides effective resistance to shear failure because the elongated intermetallic particles are able to effectively restrict the localization to a small region. Consequently, the catastrophic failure via a dominant shear band was significantly delayed since shear bands do not percolate before encountering an intermetallic network. The effective resistance by the cellular network is evidenced by the different shear directions (represented by red and yellow dashed lines in the bottom panel of Figures 8(c)-(e)) in regions separated by elongated intermetallic particles. Regular patterns formed by secondary phases have been reported to enhance plasticity in literature. For example, Hoffmann et al. [52] showed that for a $Zr_{39.6}Ti_{33.9}Nb_{7.6}Cu_{6.4}Be_{12.5}$ bulk metallic glass composite, the tensile ductility was able to reach 13% because the formation of a soft secondary phase suppressed shear band opening and inhibited crack development. In addition, Lalpoor et al. [60] observed that matching the dendrite dimension to the characteristic length scale associated with crack was essential for ductility improvement. For the present $Al_3Y$ phase, since the size of the cellular network becomes more regular with increasing hot-pressing time, the resistance to shear propagation should become higher and give rise to improved plasticity.

## 4. Discussion

In the present work, centimeter-sized Al-Mg-Y alloys with grain sizes ≤100 nm were successfully fabricated via a simple powder metallurgy approach. For $t_{HP}$ = 1 h, samples failed in a brittle fashion, while all other conditions (i.e., $t_{HP}$ = 3 h, 6 h, and 10 h) exhibited appreciable plasticity with high yield strength. The brittle versus ductile behavior is possibly due to porosity and grain sizes associated with different hot-pressing times. For $t_{HP}$ = 1 h, it is likely that samples still possess a few pores and therefore failed in a brittle fashion. This is also consistent with the lower Young's modulus for $t_{HP}$ = 1 h than for other hot-pressing times (Figure 4). While longer hot-pressing time results in improved densification, the high energy consumption associated with prolonged processing times (e.g., 10-h here) may limit practical applications of this processing route. Therefore, optimization of hot-pressing parameters, primarily hot pressing temperature and time, to achieve an excellent combination of relative density and processing efficiency is needed for potential industrial implementations. This topic represents an important direction for future applied research and technoeconomic analysis. In addition to porosity, grain sizes can strongly affect deformation behavior. Previous work suggested a transition in the deformation mechanism at a critical grain size of ~50 nm [53]. For example, Chen et al. [54] observed deformation twins in nanocrystalline Al with grain sizes of ~10-20 nm, while the coarse-grained counterpart did not show the same features due to the high stacking-fault energy. These authors further proposed a transition from partial dislocation activity dominated process to normal slip-controlled deformation as grains grow beyond the nanoscale. In the present Al-Mg-Y system, the matrix grain size is ~40 nm for $t_{HP}$ = 1 h, which is below the critical grain size of 50 nm suggested in previous studies, while grain sizes increase to ~80-110 nm for longer hot-pressing times. Therefore, the change in the deformation mechanism may also contribute to the different failure modes between short versus longer hot-pressing times.

Secondary phases can dramatically affect the mechanical behavior and failure modes since these phases usually possess different crystal structures and mechanical properties than the matrix. In the present Al-Mg-Y alloy, all hot-pressing conditions yielded the same two secondary phases ($Al_4C_3$ and $Al_3Y$), which exhibit a trigonal crystal structure [25,26] and have much higher hardness than the FCC Al matrix [55,56]. For $Al_4C_3$, the precipitate size increased with longer $t_{HP}$, while the morphology and distribution remained the same. The size of the carbides has been reported to play an important role in Al alloys, as microscale carbides have been shown to significantly deteriorate the strength of the materials [57]. In contrast, nanosized precipitates can improve both strength and plasticity [58]. Wang et al. [59] reported that nanosized $Al_6Mn$ particles at grain boundaries improve the mechanical properties of Al-6Mg-0.8Mn (wt.%). These particles have a strong pinning effect on mobile dislocations, resulting in increased dislocation densities. The accumulation of dislocations increases the driving force for dynamic recrystallization, resulting in preferential nucleation of recrystallized grains at grain boundaries. These nanosized particles also have a strong pinning effect on recrystallized grain boundaries, preventing significant grain growth and therefore contributing to the high strength of the alloy. In the present study, since the $Al_4C_3$ size is well below 1 μm for all hot-pressing conditions, these precipitates are expected to have a positive effect on both the strength and plasticity of the alloys.

In contrast, the two populations of $Al_3Y$ particles formed for all hot-pressing conditions may introduce competing effects on plasticity. When $t_{HP}$ = 1 h, equiaxed particles aggregated in regions with sizes of tens of micrometers, while larger elongated particles were located at the perimeter of these regions (Figure 2(f)). As $t_{HP}$ reached 10 h, both types of particles coarsened and the volume fraction of elongated particles dramatically increased. Furthermore, the elongated particles formed cellular networks with sizes of ~10-20 μm throughout the whole sample in Figure 2(j). Intermetallic phases usually deteriorate the plasticity of Al alloys because of their brittle nature [60]. For instance, bulk nanocrystalline $Al_{85}Ni_{10}La_5$ samples with dimensions of 2 mm × 2 mm × 4 mm failed without any plasticity under compression at a strain rate of $10^{-4}$ $s^{-1}$, which was attributed to the presence of $Al_3Ni$ and $Al_{11}La_3$ plus oxide layers and nano pores [61]. The ductility of Al alloys further decreases with increasing volume fraction of intermetallic particles. Yang et al. [62] observed shallower dimples on fracture surfaces of Al-Mg alloys after tensile testing, which corresponds to a higher fraction of $Al_3Mg_2$ phases. In addition to volume fraction, the size of intermetallic particles plays a critical role in affecting plastic deformation, with larger particles resulting in reduced plasticity. By performing quasistatic compression testing on as-cast $(Ti_{0.5}Cu_{0.23}Ni_{0.2}Sn_{0.07})_{95}Mo_5$ glass-forming alloys at room temperature, He et al. [63] observed that the plasticity decreased from ~3% to 0% when $Ti_2Ni$ particle size increased from <1 μm to a few micrometers. In contrast, the failure strain of the Al-Mg-Y system in the present study improved from ~2.8% to 6.4% when $Al_3Y$ particles coarsened from ~100-200 nm to ~500 nm-1 μm. This suggests that factors other than intermetallic particle size play a dominant role in enhancing the plasticity here. One possible explanation is the spatial distribution of the particles.

It is worth noting that the bulk compression behavior of Al-Mg-Y with $t_{HP}$ = 1 h reported here is different from that of micropillar compression performed on samples created with the same processing conditions. In contrast to the brittle failure of bulk samples, micropillars showed either stable plastic flow or localized deformation through dominant shear banding after yielding [45]. This difference can be attributed mainly to the morphology of the intermetallic particles. In Ref. [45], the cross-sectional SEM images of deformed samples showed that the vast majority of the particles had sizes below 1 μm. Although several larger intermetallic particles existed, they all exhibit a relatively equiaxed shape. However, for the bulk compression specimens in the present study, particles with a large aspect ratio formed (Figures 2 and 8), which likely result in much higher stress concentration. In Ref. [51], Pourbahari and Emamy performed tensile tests on A357 alloys with different La concentrations and observed that samples with less than 0.1 wt.% exhibited higher plasticity than samples with a higher La content. The improved plasticity in this case was attributed to intermetallic phase with a globular morphology in the low La samples compared to needle-shape intermetallic particles in the higher La content.

One key microstructural feature that contributes to the excellent plasticity of the bulk nanocrystalline Al-Mg-Y samples is the presence of amorphous grain boundary complexions.

Nanocrystalline Cu-Zr alloys with amorphous complexions were found to have much improved plasticity as compared to the same alloys with ordered grain boundary structures [64]. Moreover, Khalajhedayati et al. [64] showed that failure mode in micropillar compression became much more homogeneous when amorphous complexions were in the alloy. Wardini et al. [65] subsequently showed that improved ductility in microscale tension experiments could be achieved with these amorphous complexions. Therefore, the existence of amorphous complexions in the present Al-Mg-Y alloys can also be expected to play a significant role in enhancing plasticity because of the interactions between the endpoints of gliding dislocations and the amorphous grain boundaries.

## 4. Conclusions

In the present study, fully-dense bulk nanocrystalline Al-Mg-Y pellets were successfully fabricated using a simple powder metallurgy approach. Quasi-static bulk compression testing showed that samples with $t_{HP} > 1$ h were able to exhibit a combination of high strength and appreciable plasticity. The underlying mechanisms for high strength and plasticity were investigated thoroughly using a variety of characterization techniques. The following important conclusions are drawn:

1) All conditions resulted in the same two secondary phases, namely $Al_4C_3$ and $Al_3Y$. The $Al_4C_3$ precipitates exhibited a rod shape and remained at grain boundaries for all conditions, while two populations of $Al_3Y$ were observed: small, equiaxed versus larger, elongated particles.

2) With increasing hot-pressing time, the elongated $Al_3Y$ particles formed a cellular network which became more regular with longer pressing time. The cellular network effectively delayed shear propagation, resulting in improved plasticity for Al alloys with longer hot-pressing time.

3) The existence of amorphous complexions plays a vital role in improving the plasticity of nanocrystalline Al-Mg-Y. In addition, the retention of these complexions after naturally cooling with a slow cooling rate of less than 1 $^{o}$C/s demonstrates the excellent stability of the amorphous complexions in this alloy.

4) The hierarchy present in the microstructure of the Al alloy, including nanorod precipitates, various intermetallic particle morphology and distribution, as well as amorphous complexions, gives rise to a balance of strength and plasticity that is attractive for structural applications.

**Declaration of Competing Interest**

The authors declare that they have no known competing financial interest or personal relationships that could have appeared to influence the work reported in this paper.

**Acknowledgements**

This work was supported by the U.S. Department of Energy, Office of Science, Basic Energy Sciences, under Award No. DE-SC0025195. The authors acknowledge the use of facilities and instrumentation at the UC Irvine Materials Research Institute (IMRI), which is supported in part by the National Science Foundation through the UC Irvine Materials Research Science and Engineering Center (DMR-2011967). SEM, FIB, and EDS work were performed using instrumentation funded in part by the National Science Foundation Center for Chemistry at the Space-Time Limit (CHE-0802913).

[15] J. D. Schuler, O. K. Donaldson, and T. J. Rupert, Amorphous complexions enable a new region of high temperature stability in nanocrystalline Ni-W. *Scripta Materialia* 54 (2018) 49-53.

[16] O. K. Donaldson and T. J. Rupert, Amorphous intergranular films enable the consolidation of bulk nanocrystalline Cu-Zr with full density, *Advanced Engineering Materials* 21 (2019) 1900333.

[17] T. Lei, J. Shin, D. S. Gianola, and T. J. Rupert, Bulk nanocrystalline Al alloys with hierarchical reinforcement structures via grain boundary segregation and complexion formation, *Acta Materialia* 221 (2021) 117394.

[18] Y. M. Wang, K. Wang, D. Pan, K. Lu, K. J. Hemker, and E. Ma, Microsample tensile testing of nanocrystalline copper, *Scripta Materialia* 48 (2003) 1581-1586.

[19] E. Ma, Instabilities and ductility of nanocrystalline and ultrafine-grained metals, *Scripta Materialia* 49(7) (2003) 663-668.

[20] Z. Shan, E. A. Stach, J. M. K. Wiezorek, J. A. Knapp, D. M. Follstaedt, and S. X. Mao, Grain boundary-mediated plasticity in nanocrystalline nickel, *Science* 305 (2004) 654-657.

[21] J. A. El-Awady, Unravelling the physics of size-dependent dislocation-mediated plasticity, *Nature Communications* 6:5926 (2015)

[22] ASTM E9-19. Standard test methods of compression testing of metallic materials at room temperature.

[23] J. L .Murray, The Al-Mg (aluminum-magnesium) system, *Journal of Phase Equilibria* 3 (1982) 60-74.

[24] S. Liu, Y. Du, H. Xu, C. He, and J. C. Schuster, Experimental investigation of the Al-Y phase diagram, *J. Alloys Compd.* 414 (2006) 60-65.

[25] T. Dagerhamn and S. Westman, Note on the nine-layer hexagonal structure of YAl3, *Acta Crystallogr.* 20(6) (1966) 919-920.

[26] S. H. Li and C. G. Chao, Effects of Carbon Fiber/Al Interface on Mechanical Properties of Carbon-Fiber-Reinforced Aluminum-Matrix Composites, *Metall. Mater. Trans. A* 35A (2004) 2153-2160.

[27] I.M. Lifshitz and V. V. Slyozov, On the kinetics of oversaturated solid solution diffusion decay, *Journal of Experimental and Theoretical Physics* 35(2) (1958) 479-492.

[28] G. W. Greenwood, The growth of dispersed precipitates in solutions, *Acta Metalurgica.* 4 (1956) 243-248.

[29] M. Hillert, On the theory of normal and abnormal grain growth, *Acta Metallurgica* 13 (1965) 227-238.

[30] J. E. Burke and D. Turnbull, Recrystallization and grain growth, *Prog. Metal Phys.* 3 (1952) 220.

[31] A. L. Vasiliev, M. Aindow, M. J. Blackburn, and T. J. Watson, Phase stability and microstructure in devitrified Al-rich Al-Y-Ni alloys, *Intermetallics* 12 (2004) 349-362.

[45] T. Lei, E. C. Hessong, J. Shin, D. S. Gianola, and T. J. Rupert, Intermetallic particle heterogeneity controls shear localization in high-strength nanostructured Al alloys, *Acta Materialia* 240 (2022) 118347.

[46] Q. Wei, H. T. Zhang, B. E. Schuster, K. T. Ramesh, R. Z. Valiev, L. J. Kecskes, R. J. Dowding, L. Magness, and K. Cho, Microstructure and mechanical properties of super-strong nanocrystalline tungsten processed by high-pressure torsion, *Acta Materialia* 54(15) (2006) 4079-4089.

[47] B. Yan, Y. Zhang, S. Jiang, J. Yu, D. Sun, and M. Tang, Mechanical properties and fracture mechanisms of martensitic NiTi shape memory alloy based on various thermomechanical-processing microstructures, *Journal of Alloys and Compounds* 883 (2021) 160797.

[48] K. S. Kumar, S. Suresh, M. F. Chisholm, J. A. Horton, and P. Wang, Deformation of electrodeposited nanocrystalline nickel, *Acta Materialia* 51, 387-405 (2003).

[49] B. Schuh, R. Pippan, and A. Hohenwarter, Tailoring bimodal grain size structures in nanocrystalline compositionally complex alloys to improve ductility, *Materials Science & Engineering A* 748 (2019) 379-385.

[50] S. G. Song, N. Shi, G. T. Gray III, and J. A. Roberts, Reinforcement shape effects on the fracture behavior and ductility of particulate-reinforced 6061-Al matrix composites, *Metallurgical and Materials Transactions A* 27A (1996) 3739-3746.

[51] B. Pourbahari and M. Emamy, Effects of La intermetallics on the structure and tensile properties of thin section gravity die-cast A357 Al alloy, *Materials & Design* 94 (2016) 111-120.

[52] D. C. Hofmann, J. -Y. Suh, A. Wiest, G. Duan, M. -L. Lind, M. D. Demetriou, and W. L. Johnson, Designing metallic glass matrix composites with high toughness and tensile plasticity, *Nature* 451 (2008) 1085-1089.

[53] Y. T. Zhu and T. G. Langdon, Influence of grain size on deformation mechanisms: An extension to nanocrystalline materials, *Materials Science and Engineering A* 409(1-2) 234-242 (2005).

[54] M. Chen, E. Ma. K. J. Hemker, H. Sheng, Y. Wang, X. Cheng, Deformation twinning in nanocrystalline aluminum, *Science* 300 1275-1277 (2003).

[55] X. Tao, Y. Ouyang, H. Liu, Y. Feng, Y. Du, and Z. Jin, First-principles calculations of the thermodynamic and elastic properties of the $L1_2$-based $Al_3RE$ (Re = Sc, Y, La-Lu), *International Journal of Materials Research.* 99 (2008) 582-588.

[56] M. Besterci, L. Pešek, P. Zubko, and P. Hvizdoš, Mechanical properties of phases in $Al$-$Al_4C_3$ mechanically alloyed material measured by depth sensing indentation technique, *Materials Letters* 59 (2005) 1971-1975.

[57] D. Li, K. Hu, M. Han, K. Xie, G. Liu, and X. Liu, Strength adjust mechanism of high toughness Al4C3/Al-xMg composites, *Materials Science and Engineering A* 824 (2021) 141842.

[58] X. Li, Z. Zhang, Y. Peng, D. Yan, Z. Tan, Q. Zhou, and K. Wang, In situ synthesized nano-Al4C3 reinforced aluminum matrix composites via friction stir processing, *Journal of Materials Research and Technology* 14 (2021) 2658-2664.

[59] Y. Wang, Y. Fu, B. Yang, M. Gao, and R. Guan, Deformation behavior and microstructure evolution induced by nano-sized Al6Mn phase particles in a homogenized Al-6Mg-0.8Mn alloy during hot compression, *Materials Science and Engineering A* 882 (2023) 145461.

[60] M. Lalpoor, D. G. Eskin, G. ten Brink, and L. Katgerman, Microstructural features of intergranular brittle fracture and cold cracking in high strength aluminum alloys, *Materials Science and Engineering A* 527 (2010) 1828-1834.

[61] T. T. Sasaki, K. Hono, J. Vierke, M. Wollgarten, and J. Banhart, Bulk nanocrystalline Al85Ni10La5 alloy fabricated by spark plasma sintering of atomized amorphous powders, *Materials Science and Engineering A* 490 (2008) 343-350.

[62] B. Yang, M. Gao, Y. Wang, and R. Guan, Dynamic recrystallization behavior and mechanical properties response of rheo-extruded Al-Mg alloys with various Mg contents, *Materials Science and Engineering A* 849 (2022) 143450.

[63] G. He, J. Eckert, and W. Loser, Stability, phase transformation and deformation behvior of Ti-base metallic glass and composites, *Acta Materialia* 51(6), 1621-1631 (2003).

[64] A. Khalajhedayati, Z. Pan, and T. J. Rupert, Manipulating the interfacial structure of nanomaterials to achieve a unique combination of strength and ductility, *Nature Communications* 7(1) (2016) 10802.

[65] J. L. Wardini, C. M. Grigorian, and T. J. Rupert, Amorphous complexions alter the tensile failure of nanocrystalline Cu-Zr, *Materialia* 17 (2021) 101134.